\documentclass[12pt]{article}
\usepackage{amsmath,amsthm,amsfonts}

\newcommand{\NN}{{\mathbb N}}
\newcommand{\RR}{{\mathbb R}}

\newcommand{\dd}{\mathrm d}
\newcommand{\DD}{\mathrm D}
\newcommand{\pd}{\partial}

\newcommand{\sub}[1]{_\mathrm{#1}}
\newcommand{\I}{{\rm i}}
\newcommand{\card}{\mathop{\rm card}\nolimits}
\newcommand{\supp}{\mathop{\rm supp}\nolimits}
\newcommand{\rme}{\mathrm{e}}
\newcounter{mylc}
\renewcommand{\themylc}{\roman{mylc}}

\newcommand{\cC}{\mathcal C}
\newcommand{\cH}{\mathcal H}

\newcommand{\pp}{\mathcal S_n}
\newtheorem{theorem}{Theorem}
\newtheorem{proposition}{Proposition}
\newtheorem{corollary}{Corollary}
\newtheorem{lemma}{Lemma}
\theoremstyle{remark}
\newtheorem{remark}{Remark}
\newtheorem{conjecture}{Conjecture}
\theoremstyle{definition}
\newtheorem{definition}{Definition}
\newcommand{\etal}{et al.}

\begin{document}
\title{On quantum integrability and Hamiltonians with pure point spectrum}

\author{Alberto Enciso\thanks{aenciso@fis.ucm.es} \and Daniel Peralta-Salas\thanks{dperalta@fis.ucm.es}}
\date{\normalsize Departamento de F\'{\i}sica Te\'orica II, Facultad de
Ciencias F\'{\i}sicas, Universidad Complutense, 28040 Madrid,
Spain}
\maketitle
\begin{abstract}
We prove that any $n$-dimensional Hamiltonian operator with pure
point spectrum is completely integrable via self-adjoint first
integrals. Furthermore, we establish that given any closed set
$\Sigma\subset\RR$ there exists an integrable $n$-dimensional
Hamiltonian which realizes it as its spectrum. We develop several
applications of these results and discuss their implications in
the general framework of quantum integrability.\vspace{.3cm}

\end{abstract}
\section{Introduction}
\label{intro}
A classical Hamiltonian $h$, that is, a function from a
$2n$-dimensional phase space into the real numbers, completely
determines the dynamics of a classical system. Its complexity,
i.e., the regular or chaotic behavior of the orbits of the
Hamiltonian vector field, strongly depends upon the integrability
of the Hamiltonian.

Recall that the $n$-dimensional Hamiltonian $h$ is said to be
(Liouville) integrable when there exist $n$ functionally
independent first integrals in involution with a certain degree of
regularity. When a classical Hamiltonian is integrable, its
dynamics is not considered to be chaotic.

Given an arbitrary classical Hamiltonian there is no algorithmic
procedure to ascertain whether it is integrable or not. To our
best knowledge, the most general results on this matter are
Ziglin's theory~\cite{Zi88} and Morales--Ramis'
theory~\cite{M99}, which provide criteria to establish that a
classical Hamiltonian is not integrable via meromorphic first
integrals.

A quantum Hamiltonian $H$ is a self-adjoint linear operator acting
on the elements of a separable Hilbert space $\mathcal H$.
Proceeding by analogy with the classical case, one can define the
dimension (number of degrees of freedom) of a quantum mechanical
system~\cite{Zh89}, obtaining a notion of integrability of a
quantum Hamiltonian.

It is said that an $n$-dimensional Hamiltonian $H$ is integrable
when there exist $n$ functionally independent linear operators
$T_i$ ($i=1,\dots,n$) which commute among them and with the
Hamiltonian $H$. In Reference~\cite{Zh90} it is proved that this
definition is consistent with the classical limit in the sense
that if an integrable quantum Hamiltonian possesses a classical
counterpart, then it must be integrable as well, although the
degree of regularity of its first integrals is not specified.
However, there still exist some discrepancies with this
definition, as we will discuss in Section~\ref{final}, since this
concept does not have any geometrical content within the framework
of Quantum Mechanics.

In Reference~\cite{Zh89} it is established a criterion for quantum
integrability based on the existence of dynamical symmetries.
Unfortunately, the explicit computation of these symmetries is
usually complicated. In this paper we provide a sufficient
integrability criterion which ensures that every Hamiltonian with
pure point spectrum is integrable, allowing a spectral theoretic
approach to integrability. Furthermore, this criterion provides a
proof (and a precise statement) of a long-standing conjecture of
Percival~\cite{Pe73}.

We also manage to prove that given any closed set of real numbers,
there exists an integrable $n$-dimensional Hamiltonian which
realizes it as its spectrum. This result improves a theorem of
Crehan~\cite{Cr95}.

There exists a celebrated conjecture due to Berry~\cite{BeTa77}
which describes the statistical distribution of the point spectrum
of a quantum Hamiltonian associated with an integrable classical
Hamiltonian~\cite{Gu90} and can be stated as follows:

\begin{conjecture}[Berry]\label{conj}
The point spectrum of a generical quantum system whose
Hamiltonian yields a classically integrable system is Poisson
distributed.
\end{conjecture}

More specifically~\cite{Gu90}, the Poisson distribution
$P(s)=\rme^{-s}$ refers to the spacing
$s_i=\epsilon_{i+1}-\epsilon_i$ of the normalized energy levels
$\epsilon_i$, since the probability that $s\leq s_i\leq s+\dd s$
for a random $i$ is $P(s)\,\dd s$. This conjecture has been
recently proved for particles subjected to the action of a
magnetic field on a flat torus~\cite{Ma03}.

In this paper we will prove a closely related result ensuring that
the statistical distribution of the energy levels of a generic
quantum Hamiltonian with pure point spectrum is also Poissonian.
We also prove that for each unitary class of Hamiltonians with
pure point spectrum there exists a representative to which
Conjecture~\ref{conj} applies. This fact can be considered a
different but analogous, physically meaningful statement which
describes a purely quantum mechanical version of
Conjecture~\ref{conj} without appealing to the semiclassical
approximation and which provides additional support for Berry's
conjecture.

This paper is organized as follows. In
Section~\ref{integrabilitysection} the integrability of
Hamiltonians with pure point spectrum is studied, obtaining
additional results on the existence of integrable Hamiltonians
realizing certain prescribed spectrum in arbitrary dimension. In
Section~\ref{ber} we use this results to gain some insight into
Berry's conjecture. Finally, other interesting consequences of
this new integrability criterion are given in Section~\ref{final},
and a critical discussion of the concept of quantum integrability
is presented based on the discrepancies of its standard definition
and general wisdom, and on its lack of geometric content.

\section{Integrability of Hamiltonians with pure point spectrum}
\label{integrabilitysection}
In this section we will establish the integrability of any
$n$-dimensional Hamiltonian $H$ whose continuous spectrum is
empty. Our proof will rest upon the explicit construction of an
integrable self-adjoint operator $A$ which is completely
isospectral to our Hamiltonian $H$ in the following sense.

\begin{definition}\label{ci}
Two self-adjoint operators $A$ and $H$ are {completely
isospectral} when $\sigma\sub{cont}(A)=\sigma\sub{cont}(H)$,
$\sigma\sub{pp}(A)=\sigma\sub{pp}(H)$ and the eigenvalues of $A$
and $H$ present the same multiplicities.
\end{definition}

The definition of {point spectrum} which we will use in this
article is that of~\cite{RS72}: $\lambda$ is in
$\sigma\sub{pp}(A)$ if and only if it is an eigenvalue of the
self-adjoint operator $A$. We also use the direct sum
decomposition $\cH=\cH\sub{pp}\oplus \cH\sub{cont}$ \cite{RS72},
and define $\sigma\sub{cont}(A)=\sigma(A|_{\cH\sub{cont}})$. This
provides the decomposition
\begin{equation*}\label{decomposition}
\sigma(A)=\overline{\sigma\sub{pp}(A)}\cup\sigma\sub{cont}(A)\,,
\end{equation*}
where these two sets are not necessarily disjoint. The
self-adjoint operator $A$ will be said to have pure point spectrum
when $\sigma\sub{cont}(A)=\emptyset$. In this paper, the overline
will represent the closure of a set and $\NN_0$ will stand for the
set $\{0,1,2,\dots\}$.

Given a sequence $\cC=(E_i)_{i\in\NN_0}$ of real numbers, with
possibly repeated elements, we shall consider the associated set
\[
C=\bigcup_{i\in\NN_0}\{E_i\}
\]
of the values taken in this sequence.

\begin{definition}\label{spseq}
A self-adjoint operator $A$ is said to realize the sequence $\cC$
as its spectrum if $\sigma\sub{cont}(A)=\emptyset$,
$\sigma\sub{pp}(A)= C$ and the multiplicity of each eigenvalue $E$
of $A$ equals the times it appears in $\cC$, i.e.,
$\card\{i\in\NN_0 \;|\; E_i=E\}$.
\end{definition}

This definition clearly implies that $\sigma(A)=\overline C$. Now
we will concentrate on the construction of an integrable
Hamiltonian realizing a prescribed sequence $\cC\subset \RR$ as
its spectrum. We will follow Crehan's approach to this
problem~\cite{Cr95}.

We will need the following elementary lemma, whose proof is
straightforward and will be omitted.

\begin{lemma}\label{bijection}
Let $\cC=(E_i)_{i\in\NN_0}$ be a sequence. Then there exists a
$C^\infty$ function $f:\RR^n\to\RR$ and a bijection
$\phi:\NN_0^n\to\NN_0$ such that $f(I)=E_{\phi(I)}$ for all
$I\in\NN_0^n$.
\end{lemma}

In fact, combining the theorems of Mittag-Leffler and Weierstrass
one can prove~(\cite{Ru66}, Theorem $15.15$) that $f$ can
actually be chosen to be entire whenever $\cC$ does not possess
any accumulation points.

\begin{proposition}\label{A}
Let $\cC$ be a sequence of real numbers. Then there exists an
integrable $n$-dimensional Hamiltonian $A$, whose $n$ commuting
first integrals can be chosen to be self-adjoint, which realizes
the sequence $\cC$ as spectrum.
\end{proposition}
\begin{proof}
Let $f$ be a function as in Lemma~\ref{bijection}. Let
$N_i=\frac12(X_i^2+P_i^2-1)$ ($i=1,\dots,n$) be the number
operator associated with the $i$-th coordinate. It is clear that
these number operators commute among them: $[N_i,N_j]=0$. Let us
define $A$ by
\[
A = f(N_1,\dots,N_n)\,,
\]
for instance, via continuous functional calculus. Since the
number operators $N_i$ ($i=1,\dots, n$) are self-adjoint and
commute among them, we conclude that $A$ is also self-adjoint and
that it commutes with the number operators. These number
operators are obviously functionally independent and therefore
they constitute a complete family of commuting self-adjoint first
integrals of the Hamiltonian $A$.

The fact that these number operators act on different coordinates
also enables us to compute the spectrum of $A$ readily: its point
spectrum $\sigma\sub{pp}(A)$ is, by construction, $C$; its
continuous spectrum is empty; and the fact that $f(I)=E_{\phi(I)}$
for all $I\in\NN_0^n$, $\phi$ being a bijection, forces the
multiplicity of each eigenvalue to be given by the formula in
Definition \ref{spseq}.\end{proof}

\begin{remark}\label{clas.analog}
It is interesting to observe that every $C^\infty$ extension $f$
of the mapping $I\in\NN_0^n\mapsto E_{\phi(I)}\in C$ gives raise
to the same quantum Hamiltonian $A$. However, different choices of
this extension $f$ lead to different classical Hamiltonians via
the substitution
$a_f(x,p)=f(\frac12(x_1^2+p_1^2-1),\dots,\frac12(x_n^2+p_n^2-1))$.
Therefore we have an uncountable family of different classically
integrable $n$-dimensional Hamiltonians yielding the same
integrable quantum Hamiltonian. Note that all the orbits of these
classical Hamiltonians are bounded and generally dense on
$n$-dimensional tori.
\end{remark}
\begin{remark}
A classical Hamiltonian $h$ whose dependence on its variables is
of the form $h=h(x_1^2+p_1^2,\dots,x_n^2+p_n^2)$ is said to appear
in Birkhoff normal form. It is well known that every analytic
integrable classical Hamiltonian satisfying certain mild technical
conditions may be cast into this form~\cite{AK97}.
\end{remark}

Proposition~\ref{A} can be used to prove the existence of an
integrable Hamiltonian whose spectrum is any closed set
$\Sigma\subset \RR$. Let us recall the following lemma.

\begin{lemma}\label{dense}
Let $\Sigma\subset\RR$ be a closed set. Then there exists a
countable set $C \subset \Sigma$ which is dense in $\Sigma$.
\end{lemma}
\begin{proof}
Since complete separability is hereditary and $\RR$, endowed with
its usual metric topology, is completely separable, so is
$\Sigma$. Complete separability implies separability, so the lemma
is proved. \end{proof}

The existence of the desired Hamiltonian now stems from previous
results.

\begin{proposition}
Let $\Sigma\subset\RR$ be a closed set. Then there exists an
integrable $n$-dimensional Hamiltonian $A$ such that
$\sigma(A)=\Sigma$ and $\sigma\sub{cont}(A)=\emptyset$. Besides,
its $n$ commuting, functionally independent first integrals can be
chosen to be self-adjoint.
\end{proposition}
\begin{proof}
By Lemma~\ref{dense}, there exists a countable set
$C=\{c_i\}\subset\Sigma$ that is dense in $\Sigma$. Application of
Proposition~\ref{A} to the sequence $\cC=(c_i)$ yields the desired
result. \end{proof}

These results can be used to establish the integrability of any
Hamiltonian $H$ with pure point spectrum. Let us start proving an
auxiliary lemma.

\begin{lemma}\label{U}
Let $A$ and $H$ be two self-adjoint, completely isospectral
operators with pure point spectrum. Then they are unitarily
equivalent.
\end{lemma}
\begin{proof}
Let $\cC=(E_i)_{i\in\NN_0}$ be a sequence such that
$C=\sigma\sub{pp}(A)=\sigma\sub{pp}(H)$, each eigenvalue appearing
as many times as its multiplicity. Since $H$ is self-adjoint and
its continuous spectrum is empty, one can choose an orthonormal
basis of eigenfunctions of $H$, $\mathcal
B_H=\{{e}_i\;|\;i\in\NN_0\}$, such that $H{e}_i=E_i{e}_i$. The
same reasoning provides an orthonormal basis of eigenfunctions of
$A$, $\mathcal B_A=\{\hat e_i\;|\;i\in\NN_0\}$, such that $A\hat
e_i=E_i\hat e_i$. Set $U{e}_i=\hat e_i$ ($i\in\NN_0$) and extend
$U$ by linearity. Then $U$ is a unitary transformation and
satisfies $UH=AU$. \end{proof}

The following theorem, new in the literature, improves the results
in Crehan~\cite{Cr95} and Weigert~\cite{We92}.

\begin{theorem}\label{integrability}
Let $H$ be an $n$-dimensional Hamiltonian with pure point
spectrum. Then it is integrable and its $n$ commuting first
integrals can be chosen to be self-adjoint.
\end{theorem}
\begin{proof}
By Proposition~\ref{A}, we can construct an integrable
$n$-dimensional Hamiltonian $A$ that is completely isospectral to
$H$ and with $n$ functionally independent, commuting, self-adjoint
first integrals $N_1,\dots,N_n$. By Lemma~\ref{U}, there exists a
unitary transformation $U$ such that $H=U^\dagger AU$. Then the
operators $T_i=U^\dagger N_iU$ ($i=1,\dots,n$) constitute a
complete set of functionally independent, commuting, self-adjoint
first integrals. \end{proof}
\begin{remark}\label{dense}
The physical interest of this theorem is laid bare noting that
these operators are dense in the set of self-adjoint operators:
for every $n$-dimensional Hamiltonian $H$ there exists a family of
Hamiltonian operators $\{H_i\}_{i\in\NN}$ such that $H_i$ has pure
point spectrum (and is, therefore, integrable) and
$\|H-H_i\|\to0$. The proof is a straightforward application of the
properties of the spectral family of $H$.
\end{remark}
\begin{remark}\label{clq}
Despite the results of Zhang et al.~\cite{Zh89,Zh90}, it is not
obvious the connection between quantum and classical
integrability. We have not proved any regularity conditions of the
classical counterparts $t_i(x,p)$ of the quantum first integrals
$T_i$ ($i=1,\dots,n$), so we cannot claim that a classical
dynamical system whose quantum mechanical counterpart has pure
point spectrum must be integrable in any usual sense, i.e., via
analytic, meromorphic or smooth first integrals.

Let us consider the following example. Let $(M,g)$ be a compact
Riemannian manifold. Let $H=-\Delta$ be the Hamiltonian of a free
particle in $M$, where $\Delta$ represents the Laplace-Beltrami
operator. An appropriate choice of the domain $\DD(H)\subset
L^2(M)$ leads to a self-adjoint operator whose spectrum is known
to be discrete~\cite{Ch84} and therefore quantum integrable. The
theorem of Matveev and Topalov~\cite{MT01} on quantum
integrability of Laplacians on closed manifolds with
non-proportional geodesically equivalent metrics is thus extended
to any closed manifold using Theorem~\ref{integrability}.

However, the classical dynamical system associated to this
Hamiltonian $H=-\Delta$, whose equation of motion is just the
geodesic equation, is generally non-integrable in any usual sense.
In fact, Anosov~\cite{An67,Kl74} proved that this equation cannot
be integrable via continuous first integrals in any compact
Riemannian manifold of strictly negative sectional curvature.

It is also worth mentioning another famous example of this
phenomenon. The potential $V(x,y)=x^2y^2$ in $\RR^2$ is known to
have discrete spectrum~\cite{Si83}, so $H=-\pd_x^2-\pd_y^2+V(x,y)$
is integrable. However, its associated classical Hamiltonian
$h=p_x^2+p_y^2+V(x,y)$ is non-integrable via analytic functions as
a consequence of Yoshida's criterion~\cite{Yo87}, and in fact
numerical explorations show a complex orbit structure.

One ought to note that these examples should not be regarded as
exceptional, since in fact this will be the general case. This is
due to the fact that the unitary transformations that appear in
the quantum case do not induce symplectomorphisms of the classical
counterparts. This situation can be clearly observed in the
examples above, where the quantum Hamiltonian has been shown to be
unitarily equivalent to an integrable Hamiltonian in Birkhoff's
normal form but their classical analogues are non-integrable and
cannot be transformed into this form using a symplectomorphism.
\end{remark}

An easy, physically significant corollary can be immediately
derived from this sufficient integrability condition.

\begin{corollary}\label{coroint}
Let $H$ be an $n$-dimensional Hamiltonian. When its spectrum is
countable, it is integrable and its $n$ commuting first integrals
can be chosen to be self-adjoint.
\end{corollary}
\begin{proof}
According to Theorem~\ref{integrability} and the definition of
continuous spectrum, it is enough to prove that
$\cH\sub{cont}=\{0\}$. Recall that $\psi\in\cH\sub{cont}$ if and
only if its spectral measure $\mu_\psi$ is continuous with
respect to the Lebesgue measure~\cite{RS72}. The formula
\[
(\psi,f(H)\psi)=\int f(\lambda)\,\dd\mu_\psi(\lambda)\,,
\]applied to $f(\lambda)=1$, and the fact that $\supp\mu_\psi\subset\sigma(H)$ then combine to
show that $\|\psi\|=0$ for all $\psi\in\cH\sub{cont}$.\end{proof}

\section{Statistical distribution of Hamiltonians with pure point spectrum}\label{ber}
Let $\pp$ be the class of unitarily equivalent $n$-dimensional
Hamiltonians with pure point spectrum. Note that each element
belonging to this class is uniquely specified by its sequence of
eigenvalues $\cC$ up to unitary equivalence.
Theorem~\ref{integrability} shows that all the Hamiltonians in
this class are integrable. In Remark~\ref{clas.analog} it was
stated that for each class in $\pp$ there exists a representative
which has a (non-unique) smooth, integrable classical analogue.

Let $\cC = (\epsilon_i)_{i\in\NN_0}$ be the sequence of normalized
energies of a certain integrable Hamiltonian. The way in which
this normalization must be carried out is carefully explained
in~\cite{Gu90}. It is well known that Berry and
Tabor~\cite{BeTa77} conjectured that the statistical distribution
of the differences of normalized energies is generically
Poissonian. In Theorem~\ref{uniform} an analogous property for
standard, Hilbert-space Quantum Mechanics is proved for $\pp$.
Although clearly resembling Berry's conjecture, this theorem is of
a purely quantum mechanical nature and does not refer to any
semiclassical limit of the quantum Hamiltonian. Nevertheless, this
theorem is mathematically rigorous and physically interesting in
its own right.

It is enlightening to observe that however this theorem does
provide additional support for Berry's conjecture since
Remark~\ref{clas.analog} ensures that for each unitary class in
$\pp$ there exists a smooth, classically integrable Hamiltonian
$h$ to which Berry's conjecture applies.

Since it is readily shown that the set of normalized energy
differences is Poisson distributed when the set of normalized
energies follows the uniform distribution, as proved
in~\cite{Gu90}, our results on uniform distribution of energies
can be stated as follows.

\begin{theorem}\label{uniform}
For almost all Hamiltonians belonging to the class $\pp$, its
point spectrum is uniformly distributed.
\end{theorem}
\begin{proof}
The previous lemma implies that an element of $\pp$ is specified
by a sequence of real numbers $\cC$ up to a change of orthonormal
basis, i.e., the classes of operators in $\pp$ are in a one to one
correspondence with the sequences of real numbers. It is known
that the set of all uniform distributed sequences in a compact
space has full measure, i.e., its complement has Lebesgue measure
zero~\cite{KN74}. Although we have sequences of reals, we can
compactify $\RR$. Since the sequences in $\overline \RR$ in which
at least one element takes the value $+\infty$ or $-\infty$ have
measure zero, we reach the result that almost all (classes of)
Hamiltonians in $\pp$ verify the statement.\end{proof}

\section{Final remarks and discussion}\label{final}
In this paper we have obtained a new integrability criterion based
on the spectral properties of a quantum Hamiltonian, and proved a
result on the spectral distribution of integrable quantum
Hamiltonians which resembles a purely quantum version of Berry's
conjecture. It is important to remark that these results are
obtained in a purely quantum mechanical setting, and do not rest
upon any semiclassical treatment. It would be interesting to
develop a complete characterization of integrable self-adjoint
operators in terms of their spectrum, that is, an equivalence
between integrability and certain properties of the spectrum, and
explore the relationship between Theorem~\ref{uniform} and Berry's
conjecture.

A straightforward consequence of Theorem~\ref{integrability} is
that non-integrable quantum Hamiltonians must possess uncountable
spectra. Following the terminology of Percival~\cite{Pe73}, our
theorem implies that regular spectra (that is, spectra given by
smooth functions of the quantum numbers) can only be realized by
integrable quantum Hamiltonians since  regular spectra are always
countable. That a regular spectrum corresponds to integrability is
a long-standing conjecture of Percival for which
Theorem~\ref{integrability} provides a proof in a purely quantum
setting.

Another open question in the literature~\cite{KMS01} is the study
of Hamiltonians whose point spectrum is given by the real
solutions $c_i$ (${i\in\NN}$) of
\[
\zeta (\tfrac{1}{2}+\I c_i)=0\,, \]where $\zeta$ is the Riemann
Zeta function. Its interest is due to the fact that the statistics
associated to the energy levels spacing of $(c_i)_{i\in\NN}$ is
GUE and this statistics is generically associated to quantum
chaos. Theorem~\ref{integrability} implies that any
$n$-dimensional quantum Hamiltonian with spectrum given by
$(c_i)_{i\in\NN}$ must be integrable, improving a result of
Crehan~\cite{Cr95} which ensures the existence of an integrable
quantum Hamiltonian realizing $(c_i)_{i\in\NN}$ as its spectrum.
It should be noticed that, as already mentioned, our theorems do
not prove classical integrability, since they provide purely
quantum mechanical results. Actually some of these Hamiltonians
are known to have non-integrable classical counterparts, and
therefore quantum chaos may appear within the context of
semiclassical Quantum Mechanics.

As another nontrivial application of our integrability criterion,
we will establish the integrability of the movement of a quantum
particle in $\RR^n$ in a lower bounded potential $V$ such that
$V(x)\to+\infty$ as $x\to\infty$. Let us consider a lower bounded
potential $V\in L^r\sub{loc}(\RR^n)$ ($r>\min\{2,\frac n2\}$) such
that
\[
\lim_{R\to0}\sup_{x\in\RR^n} \int_{|x-y|\leq
R}|x-y|^{4-n-\epsilon}\, V(y)^2\,\dd^ny=0
\]for some $\epsilon>0$ and the Hamiltonian
$H=-\Delta+V(x)$, defined on
\[
\DD(H)=\big\{\psi\in L^2(\RR^n)\;\big|\;\Delta\psi\in
L^2(\RR^n)\big\}\,,
\]where the derivatives are to be understood in a distributional
sense~\cite{Br61}.

\begin{theorem}\label{P2+V}
Let the $n$-dimensional Hamiltonian $H=-\Delta +V(x)$ be defined
as above, and suppose that $V(x)\to+\infty$ as $x\to\infty$. Then
$H$ is integrable via self-adjoint first integrals.
\end{theorem}
\begin{proof}
In Reference~\cite{Br61} it is proved that $H$ is self-adjoint and
its spectrum is discrete. Hence Corollary~\ref{coroint} implies
that $H$ is integrable.\end{proof}
\begin{remark}As stated in Remark~\ref{clq}, the quantum
mechanical integrability of the Hamiltonian $H=P^2+V(X)$ in
$\RR^n$ under the aforementioned hypothesis does not imply the
integrability of its classical analogue via smooth first
integrals. It cannot be claimed that Theorem~\ref{P2+V} provides a
proof for this statement in the context of Classical Mechanics,
and actually some examples are known~\cite{SV97} in which a
potential as described above gives raise to a classical
Hamiltonian which is not integrable via meromorphic functions.
\end{remark}

We will end up with a short digression on the validity of our
results. First we should note that some authors
believe~\cite{CP90,Ku85} that Quantum Mechanics is always
completely integrable, in some sense. A simple argument goes as
follows: let $H$ be a Hamiltonian and let $\{P_\Omega\}$ be its
projection valued measure~\cite{RS72}. Then, for a generic
Hamiltonian, there exists a partition $\mathcal B$ of $\sigma(H)$
into measurable sets whose pairwise intersections have measure
zero such that $\{P_\Omega\;|\;\Omega \in\mathcal B\}$ is an
infinite family of commuting, self-adjoint, functionally
independent first integrals of $H$. Hence Quantum Mechanics is
generically superintegrable according to the definition which we
have used in this article, which is also the most popular one in
the literature.

Although we do not intend to present here a detailed study of
quantum integrability, we will point out that the results obtained
in this paper are not vacuous, since, actually, we have been
implicitly using the following stronger definition of
integrability, which in fact closely resembles the definition
proposed in~\cite{We95} in the context of (finite-dimensional)
spin dynamics.

\begin{definition}\label{inte}
An $n$-dimensional Hamiltonian $H$ is integrable when it is
unitarily equivalent to a self-adjoint operator $A$, defined on a
dense subset of $L^2(\RR^n)$, which possesses $n$ commuting,
self-adjoint, functionally independent first integrals
$N_1,\dots,N_n$ such that both $A$ and these $N_i$ are smooth
functions of the operators $(X\psi)(x)=x\psi(x)$ and $(P\psi)(x)=
-\I\nabla\psi(x)$.
\end{definition}

In more physical terms, we consider that a quantum $n$-dimensional
Hamiltonian is integrable when it can be obtained from a
canonically quantized, classically integrable Hamiltonian system
in $\RR^n$ via a change of orthonormal basis. This definition
ensures the nontriviality of our theorems.

The results of this paper arise the question of to what extent the
concept of quantum integrability can be given a non-vacuous
meaning. In Classical Mechanics this notion, when suitably
defined, merely reflects the geometric simplicity of the orbit
structure of the Hamiltonian system. However, the standard
definition of quantum integrability, which follows naively from
this classical concept, or even the slightly stronger one provided
in Definition~\ref{inte}, does not possess any geometric content,
and therefore cannot be regarded as the quantum analogue of
classical integrability. In light of these remarks, the formal
obstructions to the standard definition of quantum integrability
raised by the celebrated theorem of Von Neumann on commuting sets
of self-adjoint operators~\cite{VN96} are hardly surprising. In
this view one can also understand the amazing fact that, as stated
in Remark~\ref{dense}, integrable $n$-dimensional Hamiltonians are
dense in the set of self-adjoint operators, while it is well
known~\cite{MM74} that classically integrable Hamiltonians are
nowhere dense.

There remains as an open question to define a meaningful,
geometrically significant notion of quantum integrability, which
probably would be related to the orbit structure in the projective
Hilbert space of the quantum system and is expected to reproduce
those aspects of quantum integrability which are nowadays common
knowledge even though they are not compatible with the standard
definition of quantum integrability, establishing a clear physical
distinction between the behaviors of an integrable and a
non-integrable system. It would come to no surprise that a
geometrical definition following this philosophy were finally
independent of the number of degrees of freedom of the system,
since in fact the Hilbert spaces $\mathcal H_{N,n}$ describing the
dynamics of $N$ quantum particles of arbitrary spin moving in
$\RR^n$ are isomorphic for every choice of $N$, $n$ and the spins.
An important step towards the understanding of the concept of
quantum integrability in geometrical terms is due to Cirelli and
Pizzocchero~\cite{CP90}, but many important questions in this
field are still unanswered.

\section*{Acknowledgements}
{The authors wish to acknowledge the contribution of Professors F.
Finkel, A. Gonz\'alez-L\'opez and M.A. Rodr\'\i guez in offering
valuable suggestions during the course of this work.} AE and DPS
are supported by FPU scholarships from MECD (Spain).


\end{document}